\documentclass[12pt]{article}
\usepackage{amsmath}

\begin{document}

\begin{center}
\begin{large}
{\bf Kinetic energy properties and  weak equivalence principle in a space with GUP}
\end{large}
\end{center}

\centerline {Kh. P. Gnatenko \footnote{E-Mail address: khrystyna.gnatenko@gmail.com }, V. M. Tkachuk \footnote{E-Mail address: voltkachuk@gmail.com }}
\medskip
\centerline {\small \it Ivan Franko National University of Lviv, Department for Theoretical Physics,}
\centerline {\small \it 12 Drahomanov St., Lviv, 79005, Ukraine}

\abstract{A space with deformed commutation relations for coordinates and momenta leading to generalized uncertainty principle (GUP) is studied.
We show that GUP causes great violation of the weak equivalence principle for macroscopic bodies, violation of additivity property of the kinetic energy, dependence of the kinetic energy  on composition, great corrections to the kinetic energy of macroscopic bodies. We find that all these problems can be solved in the case  of arbitrary deformation function depending on momentum if parameter of deformation is proportional inversely to squared mass.

 }

\section{Introduction}

Investigations in String Theory and
Quantum Gravity (see, for example, \cite{Gross,Maggiore,Witten}) lead to the generalized uncertainty principle (GUP)
\begin{eqnarray}
\Delta X\geq\frac{\hbar}{2}\left(\frac{1}{\Delta P}+\beta\Delta P\right),\label{GUP}
\end{eqnarray}
here $\beta$ is a constant. From (\ref{GUP}) follows minimal position uncertainty $\Delta X_{min}=\hbar\sqrt{\beta}$. Inequality (\ref{GUP}) can be obtained from the deformed commutation relation for coordinate and momentum
\begin{eqnarray}
[X,P]=i\hbar(1+\beta P^2).\label{d}
\end{eqnarray}

In more general case the commutator for coordinate and momentum can be considered in the following form
\begin{eqnarray}
[X,P]=i\hbar F(\sqrt{\beta}|P|),\label{qgd}
\end{eqnarray}
where $F(\sqrt{\beta}|P|)$ is a deformation function, $\beta$ is the parameter of deformation, $\beta\geq0$. For $\beta=0$ the function reduces to 1 ($F(0)=1$) and relation (\ref{qgd}) corresponds to the ordinary commutation relation. To preserve the invariance of (\ref{qgd}) with respect to reflection ($X\rightarrow-X$, $P\rightarrow-P$) and  time-reversal transformation ($X\rightarrow X$, $P\rightarrow-P$, involving complex conjugation) the deformation function has to be even function (it has to be dependent on $|P|$).   Dependence on the product  $\sqrt{\beta} |P|$  is required by the dimensional considerations (from (\ref{qgd}) follows that the function is dimensionless, $\sqrt{\beta}$ has dimension of $P^{-1}$).
In the classical limit $\hbar\rightarrow0$ from (\ref{qgd}) one obtains deformed Poisson brackets
\begin{eqnarray}
\{X,P\}= F(\sqrt{\beta}|P|).\label{gd}
\end{eqnarray}
Answer on the question what functions $F(\sqrt{\beta}|P|)$ lead to the  minimal length  and explicit expression for the minimal length in
the case of arbitrary deformation function  were presented in \cite{Nowicki}.

In three-dimensional case  commutation relations for coordinates and momenta can be written as
\begin{eqnarray}
[X_i,P_j]=i\hbar {F}_{ij}(\sqrt{\beta} P_1, \sqrt{\beta} P_2, \sqrt{\beta} P_3),\label{qggd}
\end{eqnarray}
with ${F}_{ij}(\sqrt{\beta} P_1, \sqrt{\beta} P_2, \sqrt{\beta} P_3)$ being deformation functions.  For invariance of (\ref{qggd}) under time-reversal and parity transformations  it is required that functions $F_{ij}$ are even
\begin{eqnarray}
F_{ij}(-\sqrt{\beta} P_1, -\sqrt{\beta} P_2, -\sqrt{\beta} P_3)=F_{ij}(\sqrt{\beta} P_1, \sqrt{\beta} P_2, \sqrt{\beta} P_3).
  \end{eqnarray}
In the classical limit from (\ref{qggd}) one obtains deformed Poisson brackets
\begin{eqnarray}
\{X_i,P_j\}= {F}_{ij}(\sqrt{\beta} P_1,\sqrt{\beta} P_2, \sqrt{\beta} P_3).\label{ggd}
\end{eqnarray}

Different generalizations of the uncertainty principle or alternatively different deformations of commutation relations were proposed.
Among well studied algebras are the Snyder algebra (see, for instance, \cite{Snyder,Romero08,Mignemi11,Mignemi14,Lu,GnatenkoEPL19}) which in the nonrelativistic case reads
\begin{eqnarray}
[X_i,X_j]=i\hbar\beta(X_iP_j-X_jP_i),  \label{S1}\\{}
[X_i,P_j]=i\hbar(\delta_{ij}+\beta P_iP_j), \\{}
 [P_i,P_j]=0, \label{S2}
\end{eqnarray}
 deformed algebra with minimal length $\hbar\sqrt{\beta+\beta^{\prime}}$ (see, for instance, \cite{Kempf,Benczik02,Benczik05,GnatenkoIJMPD19})
 \begin{eqnarray}
 [X_i,X_j]=i\hbar \frac{(2\beta-\beta^{\prime})+(2\beta+\beta^{\prime})\beta P^2}{1+\beta P^2}(P_iX_j-P_jX_i),\label{K1}\\{}
[X_i,P_j]=i\hbar(\delta_{ij}(1+\beta P^2)+\beta^{\prime}P_iP_j),\\{}
[P_i,P_j]=0,\label{K2}
\end{eqnarray}
(here $\beta$, $\beta^{\prime}$ are constants).  Particular case of algebra (\ref{K1})-(\ref{K2}) with $\beta^{\prime}=0$  was studied (see, for instance, \cite{Kempf95,Roy1,Roy2}). For $\beta^{\prime}=2\beta$ up to the first order in $\beta$ the algebra (\ref{K1})-(\ref{K2}) transforms to algebra with commutative coordinates and commutative momenta
 \begin{eqnarray}
[X_i,X_j]=[P_i,P_j]=0,\label{C1}\\{}
[X_i,P_j]=i\hbar(\delta_{ij}(1+\beta P^2)+2\beta P_iP_j).\label{C2}
\end{eqnarray}
Also deformed algebra with $[X_i,X_j]=[P_i,P_j]=0$, and $[X_i,P_j]=i\hbar\sqrt{1+\beta P^2}(\delta_{ij}+\beta P_iP_j)$ leading to the minimal length was examined \cite{Tk3}.

 To describe a space with minimal length and maximal momentum the following deformation functions where considered ${F}_{ij}(\sqrt{\beta} P_1, \sqrt{\beta} P_2, \sqrt{\beta} P_3)=\delta_{ij}-\sqrt{\beta}(P\delta_{ij}+P_iP_j/P)+\beta(P^2\delta_{ij}+3P_iP_j)$  \cite{Ali,Ali1},  $F(\sqrt{\beta}|P|)=1/(1-\beta P^2)$ \cite{Pedram12,Pedram122}, $F(\sqrt{\beta}|P|)=(1-\sqrt{\beta}|P|)^2$ \cite{Won18}, $F(\sqrt{\beta}|P|)=1/(1-\sqrt{\beta}|P|)$ \cite{Won19}. These algebras lead to existence of minimal length and maximal momentum which are proportional to $\hbar \sqrt{\beta}$ and $1/\sqrt{\beta}$, respectively \cite{Ali,Pedram12,Won18,Won19}.

 We would like to stress that GUP leads to fundamental problems, among them are violation of the equivalence principle, violation of properties of the kinetic energy.
 These problems were examined in  the frame of algebras (\ref{d}) \cite{Tk1,Tk2}, (\ref{S1})-(\ref{S2}) \cite{GnatenkoEPL19}, (\ref{K1})-(\ref{K2})  \cite{GnatenkoIJMPD19}.  It was concluded that the way to recover the properties of the kinetic energy and to preserve the weak equivalence principle is to consider the parameter of deformation to be dependent on mass \cite{GnatenkoEPL19,GnatenkoIJMPD19,Tk1}.

In the present paper we study a space with deformation of commutation relations for coordinate and momentum  with arbitrary deformation function (\ref{gd}).
We show that the problem of violation of additivity of the kinetic energy, problem of dependence of the kinetic energy on composition, problem of grate corrections to the kinetic energy of macroscopic bodies caused by GUP and problem of violation of the weak equivalence principle can be solved in all orders in the parameter of deformation for arbitrary deformation function  $F(\sqrt{\beta}|P|)$ if one considers parameters $\sqrt{\beta}$  to be proportional inversely to mass. The results are generalized to the three-dimensional space characterized by (\ref{ggd}).

The paper is organized as follows. In Section 2 we study the properties of  kinetic energy in the frame of relations  (\ref{gd}) and (\ref{ggd}).  Section 3 is devoted to studies of influence of deformation of commutation relations for coordinates and momenta on the implementation of the weak equivalence principle.
Conclusions are presented in Section 4.

\section{Properties of kinetic energy in a space with GUP and the soccer-ball problem}

Considering Hamiltonian
$H={P^2}/{2m}$ and taking into account (\ref{gd}), one obtains the following relation between velocity and momentum
\begin{eqnarray}
\dot{X}=\frac{P}{m}{F}(\sqrt{\beta} |P|).\label{fv}
\end{eqnarray}
So, up to the first order in the parameter $\beta$ the kinetic energy can be written as
 \begin{eqnarray}
T=\frac{P^2}{2m}=\frac{m \dot X^2}{2}-{F}^{\prime}(0){\sqrt{\beta} m^2 |\dot X|\dot X^2}+(5({F}^{\prime}(0))^2-{F}^{\prime\prime}(0))\frac{\beta m^3 \dot X^4}{2}.\label{T}
\end{eqnarray}
where $F^{\prime}(x)=d F/d x$, $F^{\prime\prime}(x)=d^2 F/d x^2$.

Let us study the additivity property of kinetic energy in a space with GUP (\ref{gd}).
For this purpose we consider a system of $N$ particles with masses $m_a$ which move with the same velocities. This is equivalent to the case when a body can be divided into $N$ parts  with masses $m_a$ which can be considered as point particles.
The kinetic energy of a system (a body) is given by (\ref{T}) with $m=\sum_am_a$.
Under another consideration, according to the additivity property the kinetic energy of a system (a body) is a sum of the kinetic energies of particles forming it
 \begin{eqnarray}
T_a=\frac{m_a \dot X_a^2}{2}-{F}^{\prime}(0){\sqrt{\beta} m_a^2 |\dot X_a|\dot X_a^2}+(5({F}^{\prime}(0))^2-{F}^{\prime\prime}(0))\frac{\beta m_a^3 \dot X_a^4}{2},\label{Ta}
\end{eqnarray}
and it reads
\begin{eqnarray}
T=\sum_a T_a=\frac{m \dot X^2}{2}-{F}^{\prime}(0){\sqrt{\beta} |\dot X|\dot X^2}\sum_a m_a^2+(5({F}^{\prime}(0))^2-{F}^{\prime\prime}(0))\frac{\beta  \dot X^4}{2}\sum_a m_a^3.\label{sT}
\end{eqnarray}
Here we take into account that  $m=\sum_am_a$ and the velocity of the body and the velocities of particles forming it are the same $\dot X=\dot X_a$.

 We would like to stress that expression (\ref{sT})  does not reproduce (\ref{T}). The property of additivity of the kinetic energy is not preserved. From inequalities $m^2=(\sum_a m_a)^2>\sum_a m_a^2$ and $m^3=(\sum_a m_a)^3>\sum_a m_a^3$ follows that absolute values of first and second order corrections to the kinetic energy (\ref{T}) are bigger than absolute values of the corrections to the kinetic energy determined as (\ref{sT}). In particular case when a body (a system) is made of $N$ particles with the same masses, one has $m=Nm_a$ and expressions (\ref{T}), (\ref{sT}) read,
  \begin{eqnarray}
T=N\frac{m_a \dot X^2}{2}-N^2{F}^{\prime}(0){\sqrt{\beta} m_a^2 |\dot X|\dot X^2}+N^3(5({F}^{\prime}(0))^2-{F}^{\prime\prime}(0))\frac{\beta m_a^3 \dot X^4}{2},\label{Ti}\\
T=NT_a=N \frac{m_a \dot X^2}{2}-N\left({F}^{\prime}(0){\sqrt{\beta} m_a^2 |\dot X|\dot X^2}-(5({F}^{\prime}(0))^2-{F}^{\prime\prime}(0))\frac{\beta m_a^3 \dot X^4}{2}\right),\label{sTi}
\end{eqnarray}
respectively. Note that in expressions (\ref{Ti}), (\ref{sTi}) one has different dependence of corrections caused by GUP on the number of particles $N$. It is important to stress that from (\ref{Ti}) we have that with increasing of number of particles in a body (in a system) the corrections caused by GUP in the first and second order in $\sqrt{\beta}$ increase with $N^2$ and $N^3$ respectively, while the zero order term increases with $N$. So, according to (\ref{Ti}) influence of deformation of commutation relations (\ref{gd}) on the kinetic energy of a macroscopic system is significant.
 This problem is similar as the problem of macroscopic bodies known as  soccer-ball  problem in Double Special Relativity \cite{Hossenfelder,Amelino-Camelia,Hinterleitner,Hossenfelder1}.

It is also important to mention that according to expression  (\ref{sT}) kinetic energy of a body depends on the masses of particles forming it. For two bodies of the same masses but different compositions according to  (\ref{sT}) one obtains different kinetic energies.

 Note that the properties of kinetic energy can be preserved and the problem of grate corrections to kinetic energy of macroscopic body can be solved if we consider more general case when coordinates and momenta of different particles satisfy deformed algebra with different parameters $\beta_a$ and relate the parameters with masses $m_a$.
 In this case for a   body which can be divided into $N$ parts  with masses $m_a$ and parameters of deformation $\beta_a$ which can be considered as point particles, we can rewrite expression (\ref{sT}) as
 \begin{eqnarray}
T=\sum_a T_a=\frac{m \dot X^2}{2}-{F}^{\prime}(0){ |\dot X|\dot X^2}\sum_a\sqrt{\beta_a} m_a^2+(5({F}^{\prime}(0))^2-{F}^{\prime\prime}(0))\frac{\dot X^4}{2}\sum_a \beta_a m_a^3.\label{sTd}
\end{eqnarray}
Note that  expression (\ref{sTd}) does not depend on the masses of particles forming the body   if parameters of deformation are related with mass as
\begin{eqnarray}
\sqrt{\beta_a} m_a=\gamma=const, \label{ccc}
\end{eqnarray}
here constant $\gamma$  does not depend on mass. If relation (\ref{ccc}) holds, we can rewrite (\ref{sTd}) as
\begin{eqnarray}
T=\sum_a T_a=
\frac{m \dot X^2}{2}-{F}^{\prime}(0){\gamma m |\dot X|\dot X^2}+(5({F}^{\prime}(0))^2-{F}^{\prime\prime}(0))\frac{\gamma^2 m \dot X^4}{2}.\label{rsT}
\end{eqnarray}
Expression (\ref{rsT}) reproduces  (\ref{T})  if parameter of deformation $\beta$ corresponding to body satisfies the same condition
$\sqrt{\beta} m=\gamma$ with $m$ being  the total mass of the body.

So, if  parameter of deformation of a particle (a body) is related with its mass as
\begin{eqnarray}
\beta=\frac{\gamma^2}{m^2},\label{c}
\end{eqnarray}
 the kinetic energy in a space with GUP has additivity property and does not depend on composition.
Besides the kinetic energy (\ref{rsT}) is proportional to mass. Therefore due to relation (\ref{c}) the soccer-ball problem does not arise.

The same conclusion can be done in all orders over the parameter of deformation.
Note that if condition (\ref{c}) holds momentum $P$ is proportional to mass.
 Namely considering (\ref{c})  one can rewrite  relation (\ref{fv}) as
\begin{eqnarray}
\dot{X}=\frac{P}{m}{F}\left(\gamma \frac{|P|}{m}\right).\label{fvv}
\end{eqnarray}
 From (\ref{fvv}) follows that  $P/m$ is a function of velocity $\dot{X}$ and constant $\gamma$,
\begin{eqnarray}
\frac{P}{m}=f(\dot{X},\gamma),\label{pp}
 \end{eqnarray}\
and does not depend on mass. So, the momentum $P$ is proportional to mass $m$ as it is in the ordinary space (space with $\beta=0$).
Taking into account (\ref{pp})  kinetic energy can be written as
 \begin{eqnarray}
T=\frac{P^2}{2m}=\frac{m (f(\dot{X},\gamma))^2}{2}.\label{tt}
\end{eqnarray}
According to the additivity property the kinetic energy of a system of particles which move with the same velocities reads
 \begin{eqnarray}
T=\sum_aT_a=\sum_a\frac{m_a (f(\dot{X},\gamma))^2}{2}=\frac{m (f(\dot{X},\gamma))^2}{2},\label{add}
\end{eqnarray}
where $m=\sum_am_a$. Note that expression (\ref{add}) corresponds to (\ref{tt}) with $m=\sum_am_a$.  Note also, that the kinetic energy (\ref{add}) is proportional to the total mass of the system $m$ and does not depend on the system's composition. So, due to relation (\ref{c}) the properties of the kinetic energy are preserved.

This conclusion can be generalized to the  three-dimensional space with GUP.
Considering Hamiltonian $H=\sum_{i}P_i^2/2m$ and taking into account (\ref{ggd}) one obtains the following relation between velocities and momenta
\begin{eqnarray}
\dot X_i=\sum_j\frac{P_j}{m}{F}_{ij}(\sqrt{\beta} P_1, \sqrt{\beta} P_2, \sqrt{\beta} P_3).
\end{eqnarray}
If relation (\ref{c}) holds, we can write
\begin{eqnarray}
\dot X_i=\sum_j\frac{P_j}{m}{F}_{ij}\left(\gamma \frac{P_1}{m}, \gamma \frac{P_2}{m}, \gamma \frac{P_3}{m}\right).\label{r}
\end{eqnarray}
 From (\ref{r}) we have that $P_i/m$ are determined by velocities and constant $\gamma$,
${P_i}/{m}=f_i(\dot X_1, \dot X_2, \dot X_3, \gamma)$
 and do not depend on mass $m$. So,  the kinetic energy of a particle (a body) of mass $m$  can be rewritten as
\begin{eqnarray}
T=\sum_i\frac{m(f_i(\dot X_1, \dot X_2, \dot X_3, \gamma))^2}{2}.\label{ad0}
\end{eqnarray}
For a system (a body)  made by $N$ particles with masses $m_a$ which move with the same velocities according to the additivity property one has
\begin{eqnarray}
T=\sum_a T_a=\sum_a\sum_i\frac{m_a(f_i(\dot X_1, \dot X_2, \dot X_3, \gamma))^2}{2}=\sum_i\frac{m(f_i(\dot X_1, \dot X_2, \dot X_3, \gamma))^2}{2},\label{ad}
\end{eqnarray}
where $m=\sum_am_a$. Analyzing (\ref{ad0}), (\ref{ad}) we can conclude that the kinetic energy has the property of additivity and does not depend on composition due to relation (\ref{c}).

Note that from relation (\ref{c}) follows that particles (bodies) with different masses feel effect of GUP with different parameters.   It is important that according to (\ref{c})  there is reduction of the parameter of deformation $\beta$ of macroscopic bodies with respect to parameters of deformation of elementary particles.  For a body with mass $m$ from (\ref{c}) we obtain the following expression for parameter of deformation ${\beta}={\beta_e} m^2_e/m^2$ (here $\beta_e$ is the parameter of deformation and $m_e$  is the mass of an elementary particle). So, the parameter of macroscopic body $\beta$ is reduced by the factor $m^2_e/m^2$ with respect to parameter of deformation corresponding to an elementary particle.
 So, influence of GUP on the macroscopic bodies is less than on the elementary particles and the problem of macroscopic bodies does not arise.

 We would like also to note  that in the case of  deformed algebras leading to minimal length and maximal momentum \cite{Ali,Ali1,Pedram12,Won18,Won19}  if relation (\ref{c}) holds  minimal length  is proportional inversely to mass  $\hbar \sqrt{\beta}=\hbar\gamma/m$ and maximal momentum  is  proportional to mass $1/\sqrt{\beta}=m/\gamma$.  So, for macroscopic bodies one obtains reduction of the minimal length and increasing of the maximal momentum with respect to elementary particles.

In the next section we show  that relation of parameter of deformation with mass (\ref{c}) is also important for recovering of the weak equivalence principle in a space with GUP.

\section{Weak equivalence principle in a space with GUP}

Let us study the implementation of the weak equivalence principle in a space with GUP (\ref{gd}).
For this purpose we examine the motion of a particle with mass $m$ in  a  gravitational field $V(X)$. Considering Hamiltonian
\begin{eqnarray}
H=\frac{P^2}{2m}+mV(X)
\end{eqnarray}
and  taking into account (\ref{gd}), we find the following equations of motion
\begin{eqnarray}
\dot{X}=\frac{P}{m}{F}(\sqrt{\beta} |P|),\label{eqm1}\\
\dot{P}= -m \frac{\partial V (X)}{\partial X}{F}(\sqrt{\beta} |P|).\label{eqm2}
\end{eqnarray}
From (\ref{eqm1}), (\ref{eqm2}) one can conclude that if we consider parameter of deformation $\beta$ to be the same for different particles  the motion of a particle in  gravitational field depends on its mass and the weak equivalence principle (also known as the universality of free fall or the Galilean equivalence principle) is violated.

It is important to stress that  deformation of Poisson brackets for coordinates and momenta (\ref{gd}) leads to great violation of the weak equivalence principle. To show this let us consider the motion of two particles with masses $m_1$, $m_2$ in uniform gravitational field  and calculate the E\"otv\"os parameter.
Taking into account (\ref{eqm1}), (\ref{eqm2}) and considering  $V(X)=-gX$ (here $g$ is gravitational acceleration) up to the first order in the parameter of deformation $\beta$ we obtain
\begin{eqnarray}
\ddot{X}^{(a)}=g+3{F}^{\prime}(0)g\sqrt{\beta} m_a|\upsilon|+(2{F}^{\prime\prime}(0)-({F}^{\prime}(0))^2)g \beta m_a^2 \upsilon^2
\end{eqnarray}
here index $a$ label the particles, $\upsilon$ is velocity of particles in uniform gravitational field in the ordinary space ($\beta=0$). Therefore, up to the first order in $\beta$ the E\"otv\"os parameter reads
\begin{eqnarray}
\frac{\Delta a}{a}=\frac{2(\ddot X^{(1)}-\ddot X^{(2)})}{\ddot X^{(1)}+\ddot X^{(2)}}=3{F}^{\prime}(0)|\upsilon|\sqrt{\beta} (m_1-m_2)+(2{F}^{\prime\prime}(0)-({F}^{\prime}(0))^2)\upsilon^2\beta (m^2_1-m^2_2).\label{eta}
\end{eqnarray}
Considering $\hbar \sqrt{\beta}$ to be equal to the Planck length, $\hbar \sqrt{\beta}=l_P=\sqrt{\hbar G}/{\sqrt{c^3}}$  we can write
 \begin{eqnarray}
\frac{\Delta a}{a}=3{F}^{\prime}(0)\frac{|\upsilon|}{c}\frac{(m_1-m_2)}{m_P}+(2{F}^{\prime\prime}(0)-({F}^{\prime}(0))^2)\frac{\upsilon^2}{c^2} \frac{(m^2_1-m^2_2)}{m_P^2}\label{gv},
\end{eqnarray}
where $c$ is the speed of light and $m_P=\sqrt{\hbar c}/{\sqrt{G}}$ is the Planck mass.
 The influence of deformation on the E\"otv\"os parameter is grate. For instance, in particular case of deformation function $F(\sqrt{\beta} |P|)=1+\beta P^2$ one obtains ${F}^{\prime}(0)=0$, ${F}^{\prime\prime}(0)=2$ and for  $m_1=1$ kg, $m_2=0.1$ kg, $\upsilon=1$ m/s  one finds  $\Delta a/a\approx0.1$.
So, if we consider parameters of deformation to be the same for different particles one obtains great violation of the weak equivalence principle caused by GUP  which has to be observed experimentally. Note that tests of the weak equivalence principle shows that it holds with hight accuracy.
For instance according to results of the Lunar Laser ranging experiment the equivalence principle holds with accuracy $\Delta a/a=(-0.8\pm1.3)\cdot10^{-13}$ \cite{Williams}, the laboratory torsion-balance tests of the weak
equivalence principle give similar limits  $\Delta a/a=(0.3\pm1.8)\cdot10^{-13}$ for Be and Ti and $\Delta a/a=(-0.7\pm1.3)\cdot10^{-13}$ for Be and Al \cite{Wagner}. The MICROSCOPE space mission aims to test the validity of the weak equivalence principle  with precision $10^{-15}$ \cite{Touboul}.

It is important to note that the problem of violation of the weak equivalence principle can be solved due to condition (\ref{c}).
If relation (\ref{c}) is satisfied, $\Delta a/a$ given by (\ref{eta}) is equal to zero.
Also, if relation (\ref{c}) holds equations (\ref{eqm1}), (\ref{eqm2}) can be rewritten as
\begin{eqnarray}
\dot{X}=\frac{P}{m}{F}\left(\gamma \frac{|P|}{m}\right),\label{eqm11}\\
\frac{\dot{P}}{m}= -\frac{\partial V (X)}{\partial X}{F}\left(\gamma \frac{|P|}{m}\right).\label{eqm22}
\end{eqnarray}
Equations for $X$ and $P/m$ (\ref{eqm11}), (\ref{eqm22}) do not contain mass. Therefore, solutions of  (\ref{eqm11}), (\ref{eqm22})  $X(t)$, $P(t)/m$ do not depend on mass.
So, we can conclude that the motion of a particle in gravitational field does not depend on its mass and the weak equivalence principle is satisfied  due to relation  (\ref{c}).
This conclusion can be generalized to  three-dimensional case.

 Let us consider the Poisson brackets for coordinates and momenta  to be  given by (\ref{ggd}) and  $\{X_i,X_j\}=\{P_i,P_j\}=0$. Note that in this case the algebra is invariant with respect to translations in configurational space.
The examples of algebras of this type are algebras with ${F}_{ij}(\sqrt{\beta} P_1, \sqrt{\beta} P_2, \sqrt{\beta} P_3)=\sqrt{1+\beta P^2}(\delta_{ij}+\beta P_iP_j)$ \cite{Tk3}, ${F}_{ij}(\sqrt{\beta} P_1, \sqrt{\beta} P_2, \sqrt{\beta} P_3)=\delta_{ij}-\sqrt{\beta}(P\delta_{ij}+P_iP_j/P)+\beta(P^2\delta_{ij}+3P_iP_j)$ \cite{Ali,Ali1}, algebra given by (\ref{C1}), (\ref{C2}) \cite{Tk2,Benczik02}.

For a particle  of mass $m$ in gravitational filed $V(\bf{X})$ studying Hamiltonian
\begin{eqnarray}
H=\sum_i\frac{P^2_i}{2m}+mV(\bf{X}),\label{h}
\end{eqnarray}
and taking into account relations  (\ref{ggd}) and  $\{X_i,X_j\}=\{P_i,P_j\}=0$, one obtains the following equations of motion
\begin{eqnarray}
\dot{X}_i=\sum_i\frac{P_j}{m}{F}_{ij}(\sqrt{\beta} P_1, \sqrt{\beta} P_2, \sqrt{\beta} P_3),\label{eqm111}\\
\dot{P}_i= - m \sum_j\frac{\partial V (\bf{X})}{\partial X_j}{F}_{ij}(\sqrt{\beta} P_1, \sqrt{\beta} P_2, \sqrt{\beta} P_3).\label{eqm222}
\end{eqnarray}
Note that the motion of  particle in gravitational field does not depend on mass if  parameter of deformation is determined as (\ref{c}).  We have
\begin{eqnarray}
\dot{X}_i=\sum_j\frac{P_j}{m}{F}_{ij}\left(\gamma \frac{P_1}{m}, \gamma \frac{P_2}{m}, \gamma \frac{P_3}{m}\right),\label{eqm111c}\\
\frac{\dot{P}_i}{m}= -\sum_j\frac{\partial V (\bf{X})}{\partial X_j}{F}_{ij}\left(\gamma \frac{P_1}{m}, \gamma \frac{P_2}{m}, \gamma \frac{P_3}{m}\right).\label{eqm222c}
\end{eqnarray}
Equations for ${X}_i$, ${P}_i/m$ (\ref{eqm111c}), (\ref{eqm222c}) depend on the constant $\gamma$ which is the same for different particles and do not contain mass. So, solutions of (\ref{eqm111c}), (\ref{eqm222c}) $X_i(t)$, $P_i(t)/m$ do not depend on mass and the weak equivalence principle is recovered.

Let us study more general case when the relations of deformed algebra are given by
\begin{eqnarray}
\{X_i,X_j\}=G(P^2)(X_iP_j-X_jP_i),\label{gen1}\\
\{X_i,P_j\}=f(P^2)\delta_{ij}+F(P^2)P_iP_j,\label{gen3}\\
\{P_i,P_j\}=0.\label{gen2}
\end{eqnarray}
 From the Jacobi identity follows that the functions $G(P^2)$, $F(P^2)$, $f(P^2)$ have to satisfy relation $f(F-G)-2f^{\prime}(f+FP^2)=0$, with $f^{\prime}=\partial f/\partial P^2$ \cite{Frydryszak}. In particular case $f=1+\beta P^2$, $F=\beta^{\prime}$ one obtains deformed algebra (\ref{K1})-(\ref{K2}). The  choose $f=1$, $F=\beta$ corresponds to the nonrelativistic Snyder algebra (\ref{S1}), (\ref{S2}).

Note that from the dimensional considerations the terms $f(P^2)$ and $F(P^2)P_iP_j$ in (\ref{gen3}) have to be dimensionless therefore $f(P^2)=\tilde{f}(\beta P^2)$ and $F(P^2)=\beta \tilde F(\beta P^2)$. Also, on the basis of dimensional considerations from (\ref{gen1}) we have that $G(P^2)=\beta \tilde G(\beta P^2)$. Here $\tilde{f}(\beta P^2)$, $\tilde{F}(\beta P^2)$, $\tilde{G}(\beta P^2)$ are dimensionless functions.

Taking into account (\ref{gen1})-(\ref{gen2}), for a particle of mass $m$ with Hamiltonian (\ref{h}) one obtains the following equations of motion
\begin{eqnarray}
\dot{X}_i=\frac{P_i}{m}\tilde{f}(\beta P^2)+m\beta\sum_j\frac{\partial V (\bf{X})}{\partial X_j}\tilde G(\beta P^2)(X_iP_j-X_jP_i),\label{gen12}\\
\dot{P}_i= - m\frac{\partial V (\bf{X})}{\partial X_i}\tilde{f}(\beta P^2)-m\beta\sum_j \frac{\partial V (\bf{X})}{\partial X_j}\tilde F(\beta P^2)P_iP_j.\label{gen13}
\end{eqnarray}
Note, that if relation (\ref{c}) is satisfied one can write
\begin{eqnarray}
\dot{X}_i=\frac{P_i}{m}\tilde{f}\left(\gamma^2 \frac{P^2}{m^2}\right)+\gamma^2\sum_j\frac{\partial V (\bf{X})}{\partial X_j}\tilde G\left(\gamma^2 \frac{P^2}{m^2}\right)\left(X_i\frac{P_j}{m}-X_j\frac{P_i}{m}\right),\label{gen112}\\
\frac{\dot{P}_i}{m}= -\frac{\partial V (\bf{X})}{\partial X_i}\tilde{f}\left(\gamma^2 \frac{P^2}{m^2}\right)-\gamma^2\sum_j \frac{\partial V (\bf{X})}{\partial X_j}\tilde F\left(\gamma^2 \frac{P^2}{m^2}\right)\frac{P_iP_j}{m^2}.\label{gen113}
\end{eqnarray}
 The solutions of (\ref{gen112}), (\ref{gen113}) ${X}_i(t)$, ${P}_i(t)/m$ do not depend on mass. Therefore due to condition (\ref{c}) the motion of a particle in gravitational field in the space (\ref{gen1})-(\ref{gen2}) does not depend on its mass and the weak equivalence principle is recovered.

\section{Conclusion}
 A space characterized by deformation of commutation relations  for coordinates and momenta with arbitrary deformation function  (\ref{gd}) has been studied.
 The properties of the kinetic energy of composite system (macroscopic body), weak equivalence principle have been examined in the space (\ref{gd}).

 We have shown that considering parameter of deformation $\beta$ to be the same for different particles (bodies) one faces the problem of violation of  additivity of kinetic energy  and its dependence on composition.
  Besides  if parameter of deformation is the same for different particles one obtains grate corrections to the kinetic energy of a composite system caused by deformation of commutation relations (\ref{gd}).  Namely we have shown that the corrections in the first and second orders in $\sqrt{\beta}$ increase with $N^2$ and $N^3$ respectively, with increasing of the number of particles $N$ in the system (\ref{Ti}). So, one faces the problem of macroscopic bodies (soccer-ball problem) in a space with GUP.
 We have also concluded that the deformation of commutation relation (\ref{gd}) leads to grate violation of the weak equivalence principle (\ref{eta}) which has to be observed at the experiment.

 We have found that if parameter of deformation depends on mass as (\ref{c})   the weak equivalence principle is recovered and the properties of the kinetic energy are preserved in all orders in the parameter of deformation in a space characterized by deformed commutation relations (\ref{gd})  with arbitrary function of deformation. This conclusion has been generalized to the three-dimensional space characterized by (\ref{ggd}).

It is worth noting that idea to relate parameters of  algebra for coordinates and momenta with mass is also important in noncommutative phase space of canonical type \cite{GnatenkoPLA13,GnatenkoPLA17}, in a  Lie-algebraic noncommutative space \cite{GnatenkoPRD19}. In the present paper we have shown that this idea is important in the frame of deformed algebras (\ref{gd}) with arbitrary deformation functions. The number of algebras and importance of results which can be obtained due to this idea justify its significance.

\section*{Acknowledgements}

The publication contains the results of studies conducted by President's of Ukraine
grant for competitive projects (F-82, No. 0119U103196). This work was partly supported by the Projects $\Phi\Phi$-83$\Phi$
(No. 0119U002203) and  $\Phi\Phi$-63Hp
(No. 0117U007190) from the Ministry of Education
and Science of Ukraine.

\end{document}